\documentclass[manuscript]{aastex}
\begin{document}

\title{EMISSIVE MECHANSIM OF RADIO FLAT SPECTRUM ON X-RAY BINARIES}
\author{Jiancheng Wang}

\affil{Yunnan Observatory, Chinese Academy of Sciences, P.O. Box 110,
Kunming, Yunnan Province, 650011, P.R. China} \affil{National Astronomical
Observatories, Chinese Academy of Sciences} \affil{United Laboratory of
Optical Astronomy, Chinese Academy of Sciences} %
\authoremail{jcwang@public.km.yn.cn}

\begin{abstract}
We present that the radio emission with flat spectrum in X-ray binaries comes from the
synchrotron emission of relativistic electrons in the high energy tail of
hot electrons in continuous conical jet. The jet is assumed to be produced by the
advection-dominated accretion flow (ADAF) and maintains ion and electron
temperatures constant in the case of adiabatic steady conical expansion.
The flat spectrum is result of self-absorbed synchrotron emission by
relativistic thermal electrons. We find that the critical frequency at which the
radiation becomes optically thin declines along the jet. The emission
observed at higher frequencies originates at smaller distance, closer to
the base of the jet. The highest cut-off frequency of the flat spectrum
is at the base of the jet, and is determined by the physics of the ADAF and
the position of the jet formation. We assert that it is a characteristic
of the ADAF in black hole X-ray binaries that a continuous steady outflow
is formed and causes the observed flat spectrum in the low/hard state.
The observed synchrotron emission consists of the flat spectral component
from the jet and the steep spectral component from the ADAF. The flat
spectral component extends from infrared to radio wavelengths,
while the steep spectral component with the 2/5 spectral slope extends from
infrared to shorter wavelengths, it will be dominated by the thermal
emission from companion star.
\end{abstract}

\keywords{accretion, accretion disks: stars---binaries: close---ISM:
jets and outflows---radiation mechanisms---radio continuum: stars}

\section{INTRODUCTION}

Radio emission is observed from 20\% of X-ray binaries (e.g. Hiellming \&
Han 1995). In many cases of persistent or repetitive radio emission, there
appears to be an underlying flat-spectrum component in low/Hard X-ray state
(Fender 2000 and references therein). This spectral component is usually
considered to originate in partially self-absorbed synchrotron emission from
conical jets, of same genre as those originally considered for Active
Galactic Nucleus (AGNs) (Marscher \& Gear 1985; Hjellming \& Johnston 1988;
Falcke \& Biermann 1996, 1999). These models assume the nonthermal electrons
with a power-law distribution to produce a flat spectrum from the jet with a
variety of geometries, magnetic fields and energetics, and produce only a flat
spectrum over at most three decades in frequency. The radio spectrum
observed from these black hole systems in the low/Hard state is quite
different, showing a flat spectrum which probably extends to very high
frequencies. For example, the flat radio spectrum may extend to mm
wavelengths for Cyg X-1 (Fender et al. 2000) and even to near-infrared bands
for GX 339-4 (Fender 2001) and GRS 1915+105 (Fender et al. 1997; Mirabel et
al. 1998; Fender \& Pooley 1998, 2000). It is therefore clear that X-ray
binaries have much flatter radio-mm (-infrared) spectra than the
'flat-spectrum' AGNs. They may be not relevant in jet kinematics and
emissive mechanisms. The model of optically thin emission from a flatten
electron spectrum (Wang et al. 1997) is not applied to X-ray binaries. The
frequency-dependent time delays observed in GRS 1915+105 (Fender et al.
1997; Fender \& Pooley 1998) argue against this model, as they imply
significant optical depth. Another possible mechanisms are a optically thin
free-free emission and a combination of free-free emission and synchrotron
emission (Fender et al. 2000), but the detailed calculations are needed.
Therefore the model suitable for X-ray binaries remains to be estimated.

The extremely strong correlation between the hard X-rays and the radio
emission in many X-ray binaries (e.g. GX 339-4 [Fender et al. 1999; Corbel
et al. 2000], GRO J1655-40 [Harmon et al. 1995], GRS 1915+105 [Harmon et al.
1997; Fender et al. 1999], Cyg X-3 [McCollough et al. 1999], and Cyg X-1
[Brocksopp et al. 1999]) has been observed. They imply that the regions
responsible for emission in the two energy regimes are strongly physically
coupled. It is currently believed that the hard X-ray emission is produced
by hot electrons via comptonisation, it seems likely that the high-energy
electrons responsible for the radio emission are simply the high-energy tail
of hot electrons.

In this letter we present that the flat spectra of the low/hard state
components are produced by the high-energy tail of hot electrons in
continuous conical jets. The jets are considered to arise in the
advection-dominated accretion flow (ADAF) (Narayan \& Yi 1995; Esin et al.
1998) or the advection-dominated inflow-outflow (ADIO) (Blandford \&
Begelman 1999). We study the jet kinetics and show that the hot electrons
maintain constant temperature in the adiabatic steady expansion of the
conical jet. The relativistic thermal electrons in the jet turn out to
produce the flat spectrum by self-absorbed synchrotron emission.

\section{JET KINEMATICS AND EMISSIVE MECHANISM}

Assume that the jet is a one-dimensional adiabatic steady conical, which
starts at the distance $z_0$ to the central black hole, with constant
opening half-angle $\phi $ and constant velocity $u$. The jet follows the
continuity of momentum $(p+\rho u^2)$, the energy $(\frac 12\rho u^3+\rho
uw) $, and the mass $(\rho u)$ flux densities, e.g.,
\begin{equation}
\rho uA(z)=const,\quad (p+\rho u^2)A(z)=const,\quad (\frac 12\rho u^3+\rho
uw)=const,
\end{equation}
where $A(z)=\pi (tg\phi )^2z^2$ is the cross-section of the jet at the
distance $z$, then the thermodynamic quantities of the jet as function of
distance $z$ have simple form: $\rho =\rho _0\left( z/z_0\right) ^{-2}$, $%
p=p_0(z/z_0)^{-2}$, and $w=w_0(z/z_0)^{-2}$.

We assume that the gas in the jet from ADAF is equipartition with an
isotropically tangled magnetic filed. We take $\beta $ to be independent of $%
z$, and let the total pressure $p$ as
\begin{equation}
p=p_g+p_m,\quad p_g=\beta p,\quad p_m=(1-\beta )p,
\end{equation}
where $p_g$ is the gas pressure, $p_m$ is the magnetic pressure. For the jet
plasma from ADAF, we allow the ion temperature $T_i$ and the electron
temperature $T_e$ to be different, and take the gas pressure to be given by
(Narayan \& Yi 1995)
\begin{equation}
p_g=\frac{\rho k}{m_u}\left( \frac{T_i}{\mu _i}+\frac{T_e}{\mu _e}\right)
=6.72\times 10^7\rho (T_i+1.08T_e)=\beta p,
\end{equation}
where the effective molecular weights of the ions and electrons are $\mu
_i=1.23$ and $\mu _e=1.14$ which correspond to a hydrogen mass fraction $%
0.75 $. The magnetic pressure is defined as $p_m=\frac{B^2}{8\pi }=(1-\beta
)p$. With the above relations, we find the distribution of physical
quantities along the jet as follows:
\begin{equation}
\rho =\rho _0\left( \frac z{z_0}\right) ^{-2},\quad B=B_0\left( \frac
z{z_0}\right) ^{-1},\quad T_i+1.08T_e=const.
\end{equation}
It is shown that the ion and electron temperatures remain constant along the
jet. The temperatures of ions and electrons in ADAF are very high. The ions
are close to their viral temperature $10^{12}K$, while the electrons are at $%
10^9K$. Therefore the electrons in the jet are also at higher temperature $%
10^9K$. Due to the assumption an equipartition magnetic field in the jet
plasma, synchrotron emission from relativistic electrons in the high energy
tail of Maxwellian distribution provides likely the observed flat radio
emission. We now turn to take the detailed calculation of synchrotron
emission from relativistic thermal electrons. The emissive coefficient of
synchrotron radiation by relativistic thermal electrons is given by
(Narayan \& Yi 1995; Mahadevan, Narayan, \& Yi 1996; Mahadevan 1997)
\begin{equation}
\epsilon _s=4.43\times 10^{-30}\frac{4\pi n_e\nu }{K_2(1/\theta _e)}M(x_M),
\end{equation}
where $M(x_M)$ is given by
\begin{equation}
M(x_M)=\frac{4.0505}{x_M^{1/6}}\left( 1+\frac{0.40}{x_M^{1/4}}+\frac{0.5316}{%
x_M^{1/2}}\right) \exp (-1.8899x_M^{1/3}),
\end{equation}
and
\begin{equation}
x_M\equiv \frac{2\nu }{3\nu _b\vartheta _e^2},\quad \nu _b\equiv \frac{eB}{%
2\pi m_ec},\quad \theta _e=\frac{kT_e}{m_ec^2}
\end{equation}
The synchrotron photons in the jet plasma are self-absorbed and give a
blackbody spectrum, up to a critical frequency $\nu _c$. The frequency at
which this occurs, at the distance $z$, is determined by
\begin{equation}
2\pi \frac{\nu _0^2}{c^2}kT_eA(z_0)+\int_{z_0}^z\varepsilon _sA(z)dz=2\pi 
\frac{\nu _c^2}{c^2}kT_eA(z),
\end{equation}
where the first term in the left side of Eq(8) is the synchrotron emission
getting into the jet at the distance $z_0$, the second term is the
synchrotron emission over a volume of conical jet between $z_0$ and $z$, and
the first term in the right side of Eq(8) is the synchrotron emission
through the cross-section at the distance $z$. We assume the emissive
coefficient $\epsilon _s$ to be constant, we obtain the relation from Eq(8)
\begin{equation}
x_M(z)=6.15\times 10^{-11}\frac{4\pi n_ez}B\frac 1{\theta _e^3K_2(1/\theta
_e)}M[x_M(z)].
\end{equation}
From Eq(4) we find that $4\pi n_ez/B$ remain constant along the jet, and
that $x_M$ is also constant in the jet. Given $x_M$, the cutoff frequency at
the distance $z$ is given by
\begin{equation}
\nu _c=\frac 32\theta _e^2\nu _bx_M=\nu _0\left( \frac z{z_0}\right) ^{-1}
\end{equation}
where $\nu _0=\frac{3eB_0\theta _e^2x_M}{4\pi m_ec}$ is the cutoff frequency
at the base of the jet, and $A(z)\propto \nu _c^{-2}$. At the frequency $\nu
_c$, the radiation becomes optically thin, and the luminosity is given by
the Rayleigh-Jeans part of the blackbody spectrum ($h\nu /kT_e\ll 1$)
\begin{equation}
L_{\nu _c}=2\pi \frac{\nu _c^2}{c^2}kT_eA(z)=2\pi \frac{\nu _0^2}{c^2}%
kT_eA(z_0)
\end{equation}
which shows that the luminosity at the frequency $\nu _c$ is independent of
the frequency $\nu _c$. This produces a flat spectrum, which extends from $%
\nu _0$ down to $\nu _{\min }$, where $\nu _{\min }$ is the cutoff frequency
given by setting $z=z_{\max }$ in Eq(10). Beyond this distance, the jet
shrinks and causes a steeper radio spectrum, as long as the electron
temperature declines (below $T_e=10^8K$, there is no synchrotron radiation).
Eq(9) shows how the cutoff frequency varies with $z$. Emission observed at
higher frequencies originates at smaller distance, closer to the base of the
jet. The peak frequency is at the base of the jet.

We can use the self-similar solutions of ADAF disk to estimate the peak
frequency $\nu _0$ and the total luminosity $\nu L_\nu $ of a flat-spectrum
source. At the base of the jet, the magnetic field is given by (Narayan \&
Yi 1995)
\begin{equation}
B_0=2.22\times 10^5m^{-1/2}\stackrel{.}{m}^{1/2}h_1\left( \frac{z_0}{10^3R_s}%
\right) ^{-5/4} G,
\end{equation}
and $h_1=(\alpha /0.3)^{-1/2}[(1-\beta
)/0.5]^{1/2}(c_1/0.5)^{-1/2}(c_3/0.3)^{1/4}$, where we scale the mass of
central object in solar unit by writing $M=mM_{\odot }$ and accretion rate
in Eddington unit, $\stackrel{.}{M}=\stackrel{.}{m}\stackrel{.}{M}_{Edd}$, $%
R_s=2GM/c^2=2.95\times 10^5m$ cm, $\alpha $ is the standard viscosity
parameter (Shakura \& Sunyaev 1973), and $c_1$, $c_3$ are constants as
defined in Narayan \& Yi (1995). With the Eq(10) and Eq(11), the peak
frequency and total luminosity are given by
\begin{equation}
\nu _0=2.68\times 10^{13}m^{-1/2}\stackrel{.}{m}^{1/2}h_1\left( \frac{x_M}{%
10^3}\right) \left( \frac{T_e}{10^9K}\right) ^2\left( \frac{z_0}{10^3R_s}%
\right) ^{-5/4}Hz,
\end{equation}
\begin{equation}
\nu L_\nu =1.61\times 10^{30}m^{1/2}\stackrel{.}{m}^{3/2}h_1^3\Omega \left( 
\frac{x_M}{10^3}\right) ^3\left( \frac{T_e}{10^9K}\right) ^7\left( \frac{z_0%
}{10^3R_s}\right) ^{-7/4}ergs~s^{-1}
\end{equation}
where $\Omega =\pi (tg\phi )^2$, and the parameter $x_M$ satisfies the Eq(9)
given by
\begin{equation}
x_M=1.12\times 10^{10}(m)^{1/2}(\stackrel{.}{m})^{1/2}\left( \frac{z_0}{%
10^3R_s}\right) ^{3/4}h_2\frac 1{\theta _e^3K_2(1/\theta _e)}M(x_M),
\end{equation}
and $h_2=\{(\alpha /0.3)[(1-\beta )/0.5)](c_1/0.5)(c_3/0.3)\}^{-1/2}$. Table
1 shows the values of $\nu _0$ and $\nu L_\nu $ for a set of temperatures $%
T_e$ at given $z_0=10^3R_s$, $m=10$ and $\stackrel{.}{m}=10^{-2}$, in which
we have used the values of $\theta _e^3K_2(1/\theta _e)$ given by Mahadevan
(1997).

We now estimate the total internal jet power. The power is given by
\begin{equation}
L_{jet}=(\frac 12\rho u^3+\rho uw)A(z_0), 
\end{equation}
where the enthalpy $w$ can be writes as $w=\frac \gamma {\gamma -1}\frac
p\rho =\frac \gamma {\gamma -1}c_s^2$, $\gamma $ is adiabatic index and $c_s$
is isothermal sound speed. We take $\gamma =5/3$ and obtain $w=2.5c_s^2$.
Assume $u<\sqrt{2w}$, e.g.,
\begin{equation}
u<\sqrt{5}c_s(z_0)=2.74\times 10^{-2}c\left( \frac{c_3}{0.3}\right)
^{1/2}\left( \frac{z_0}{10^3R_s}\right) ^{-1/2}, 
\end{equation}
where $c$ is the speed of light, we obtain the jet power as
\begin{equation}
L_{jet}\simeq \frac 52\rho uc_s^2A(z_0)=1.30\times 10^{34}m\stackrel{.}{m}%
h_3\Omega \left( \frac u{10^{-3}c}\right) \left( \frac{z_0}{10^3R_s}\right)
^{-1/2}ergs~s^{-1}, 
\end{equation}
where $h_3=(\alpha /0.3)^{-1}(c_1/0.5)^{-1}(c_3/0.3)^{1/2}$.

For a given accretion rate and matter-to-energy conversion ($\eta _{eff}=0.1$%
), standard accretion disks predict a total accretion luminosity is $%
L_{disk}=\stackrel{.}{m}c^2\stackrel{.}{M}_{Edd}=1.25\times 10^{39}m%
\stackrel{.}{m}ergs~s^{-1}$. The ADAF produces a lower luminosity
because most of the viscously dissipated energy is advected inward with the
flow and deposited into the black hole instead of being radiated. For $%
\stackrel{.}{m}>10^{-3}\alpha ^2$, the total luminosity of the ADAF is given
by (Mahadevan 1997)
\begin{equation}
L_{ADAF}\simeq 2.70\times 10^{38}m\stackrel{.}{m}^2h_4\quad ergs~s^{-1}, 
\end{equation}
where $h_4=g(\theta _e)(\alpha /0.3)^{-2}(c_3/0.3)(\beta /0.5)(r_{\min
}/3)^{-1}$, and $0.5<g(\theta _e)<13$ for temperature ranges of interest
(see Mahadevan 1997).

The thermal self-absorbed synchrotron radiation from the ADAF has a spectrum
with 2/5 spectral index which extends from $\nu _p$ down to $\nu _0$, where $%
\nu _p$ is the cutoff frequency given by setting $r=r_{\min }=3$ in equation
(13) (Mahadevan et al. 1997)
\begin{equation}
\nu _p=2.85\times 10^{15}m^{-1/2}\stackrel{.}{m}^{1/2}h_1\left( \frac{x_M}{%
10^3}\right) \left( \frac{T_e}{10^9K}\right) ^2Hz 
\end{equation}

Therefore the observed synchrotron emission consists of the flat spectral
component from the jet and the steep spectral component from the ADAF. Since
the steep spectral component extends from infrared $\nu _0$ to shorter
wavelengths $\nu _p$, it will be extremely hard to detect, being more weaker
than thermal emission from companion star.

\section{DISCUSSION AND CONCLUSIONS}

We have presented that the radio emission with flat spectrum comes from the
synchrotron emission of relativistic electrons in the high energy tail of
hot electrons in continuous conical jets. The jet is produced by the
ADAF/ADIO and maintains ion and electron temperatures constant in the case
of adiabatic steady conical expansion. The flat spectrum is result of
self-absorbed synchrotron emission by relativistic thermal electrons. The
critical frequency at which the radiation becomes optically thin declines
with the distance $z$. The emission observed at higher frequencies
originates at smaller distance, closer to the base of the jet. The highest
cut-off frequency of the flat spectrum is at the base of the jet, and is
determined by the physics of ADAF and the position of the jet formation. We
assert that it is a characteristic of the ADAF in black hole X-ray binaries
that a continuous steady outflow is formed and causes the observed flat
spectrum in the low/hard state. In the high/soft state, the hot thin
accretion disk may extend much closer to the black hole, the outflow is not
present. The radio emission is suppressed with respect to the low/hard
state. In the transition state, the ADAF may be at the stage of growth or
shrink and produce discrete ejection blobs. The ejected blob will expand
rapidly and strongly interact with its surrounding to cause blast waves. The
blast waves then accelerate a group of thermal electrons to be relativistic
and produce optically thin radio emission.

The simultaneous radio-infrared oscillations have been observed in GRS
1915+105 (Fender et al. 1997; Fender \& Pooley 1998), the infrared-radio
delay, as well as delays within the radio band, occurs. If a disturbance
travels along the jet and cause the changes of physical parameters, the
infrared-radio oscillations will be appeared. The delay between high and low
frequency emissions is the natural result of self-absorbed synchrotron
emission occurred at the different region of the jet. A constant electron
temperature in the jet remains within the distance $r_{\max }$. Beyond this
distance, the electron temperature declines as the jet shrinks, which
produces a steeper radio spectra. It can explain the observed results of
four systems, GS 2023+338, GRO J0422+32, GS 1354-64 and GRS 1915+105 (Fender
2001 and references therein).

It also is a characteristic of ADAF that the strong correlation between the
presence of hard X-ray emission and radio emission observed in persistent
black hole candidate X-ray binaries. Since the relativistic thermal
electrons responsible for the radio emission in the jet arise in ADAF and
directly indicates a population of hot electrons. The hard X-ray emission
arises in inverse Comptonisation of soft photons by hot electrons in the
ADAF and the jet. The ADAF will be significant enough to contribute in hard
X-rays if the jet carries a small part of hot electrons in the ADAF. The
hard X-ray spectrum is a power-law which extends to $\nu
=3kT_e/h=260keV(T_e/10^9K)$. The spectral index is determined by the optical
depth to electron scattering and amplification factor of photon energy in
one scattering (Mahadevan 1997). In the high/soft state the ADAF is not
present, the hard X-ray emission is suppressed. If the jet has larger
velocity, it will be a significant power output channel for the ADAF and
affect the evolution of the ADAF in the low/hard and transition state.

The observed synchrotron emission consists of the flat spectral component
from the jet and the steep spectral component from the ADAF. The flat
spectral component extends from infrared $\nu _0$ to radio wavelengths,
while the steep spectral component with the 2/5 spectral slope extends from
infrared $\nu _0$ to shorter wavelengths $\nu _p$, it will be dominated by
the thermal emission from companion star.

\clearpage
\begin{table}
\begin{center}
\caption{The peak frequency and total luminosity for the electron temperature
range of interest. \label{tbl-1}}
\begin{tabular}{cccc}
\tableline\tableline
$T_e$ & $x_M$ & $\nu _0$ & $\nu L_\nu $ \\ 
$10^9K$ & $10^3$ & $10^{12}Hz$ & $10^{29}ergs~s^{-1}$ \\
\tableline
1.00 & 3.230 & 2.74E+00 & 1.72E+00 \\ 
1.50 & 2.269 & 4.32E+00 & 1.02E+01 \\ 
2.00 & 1.800 & 6.10E+00 & 3.80E+01 \\ 
2.50 & 1.516 & 8.03E+00 & 1.08E+02 \\ 
3.00 & 1.321 & 1.01E+01 & 2.56E+02 \\ 
3.50 & 1.177 & 1.22E+01 & 5.34E+02 \\ 
4.00 & 1.065 & 1.44E+01 & 1.01E+03 \\ 
4.50 & 0.975 & 1.67E+01 & 1.76E+03 \\ 
5.00 & 0.901 & 1.91E+01 & 2.91E+03 \\ 
5.50 & 0.838 & 2.15E+01 & 4.56E+03 \\ 
6.00 & 0.784 & 2.39E+01 & 6.88E+03 \\ 
6.50 & 0.738 & 2.64E+01 & 1.00E+04 \\ 
7.00 & 0.696 & 2.89E+01 & 1.20E+04 \\ 
7.50 & 0.660 & 3.14E+01 & 1.95E+04 \\ 
8.00 & 0.627 & 3.40E+01 & 2.63E+04 \\ 
8.50 & 0.598 & 3.66E+01 & 3.48E+04 \\ 
9.00 & 0.571 & 3.92E+01 & 4.53E+04 \\ 
9.50 & 0.546 & 4.17E+01 & 5.79E+04 \\ 
10.0 & 0.524 & 4.44E+01 & 7.32E+04 \\
\tableline
\end{tabular}
\end{center}
\end{table}

\end{document}